# Decomposition and Analysis of Technological domains for better understanding of Technological Structure

*Xin Guo[a], Hyunseok Park[b], Christopher L. Magee[c]*

[a] Department of Applied Mathematics, ENSTA ParisTech, Paris, France
[b] Postdoctoral Associate, Massachusetts Institute of Technology, Cambridge, MA
[c] Professor, Institute for Data, Systems, and Society, Massachusetts Institute of Technology, Cambridge, MA

*Student: xin.guo@ensta-paristech.fr*
*Mentors: parkhs@mit.edu, cmagee@mit.edu*

**ABSTRACT**
Patents represent one of the most complete sources of information related to technological change. This paper presents three months of research on U.S. patents in the field of patent analysis. The methodology consists of using search terms to locate the most representative international and US patent classes and determines the overlap of those classes to arrive at the final set of patents and using the prediction model developed by Benson and Magee to calculate the technological improvement rate for the technological domains. My research focused on the Biochemical Pharmacology technological area and selecting relevant patents for technological domains and sub-domains within this area. The goal is to better understand structure of technology domain and understand how fast the domains and their sub-domains progress. The method I used is developed by Benson and Magee which is called the Classification Overlap Method1, it provides a reliable and largely automated way to break the patent database into understandable technological domains where progress can be measured.

**INTRODUCTION**
Christopher L. Magee, Professor of the Practice of Engineering System, Co-Director, International Design Center, Singapore University of Technology & Design and MIT and his PhD student Christopher L. Benson (who has now graduated), are investigating in the field of innovation and technology progress. They are trying to understand why certain technological domains such as information technology achieve high improvement rates while others such as batteries improve much more slowly. Professor C.L. Magee and Chris Benson studied technological capabilities in 28 technological domains and demonstrated technological capabilities have been growing exponentially with time, which can be viewed as generalization for Moore's law.

*Research Goal*

Based on the previous research, we want to look into a deeper level of technological domains and to better understand the technological structure. A technology area may include several technological domains and one technological domain may also include many different specific sub-domains. How we understand the technological performance change depends on how we define technological domains. The goal is to decompose technological domains to sub-domains and compare their technological performance change.

*Definition of technological domain and Sub-domain*

A technology domain is a set of artifacts that fulfill a generic function using a specified body of technical knowledge2. A technological sub-domain is a deeper level domain which is included in the technological domain. For example if we define 'Batteries' as a technological domain as it fulfills the generic function: Energy Storage. Then we can define several different types of battery as its sub-domains. Many areas of academic and industrial work make use of the notion of a 'technology'. As technological progress continues to accelerate, a greater need arises to understand how technologies advance over time and how we can implement those lessons into developing technologies for the future.
One of the sources of data that has been widely used for understanding technological growth is the patent data that approximately records most of the advances in technology.
Professor Christopher and Benson has developed a method called classification overlap method (com) which provides a reliable and largely automated way to break the patent database into understandable technological domains where progress can be measured.
As a technological domain may include several deeper level sub-domains and enterprise may be more interested in the technological
progress of a more specific technology, we want to decompose the technological domain into sub-domains and see how they progress.

*Calculation of technological performance change rate (k-value)*

It is possible to quantify the improvement of a technological domain over time, as first introduced by Moore and has since been explored more broadly and deeply by many others. All of these authors find exponential relationships between performance and time or equivalently that the fractional (or percentage) change per year is constant. Specifically, if q is performance at time t and $q_0$ performance at a reference time, $t_0$,

$$q = q_0 \exp(k(t - t_0))$$

The exponential constant k is referred to here as the technological improvement rate.
The technological improvement rate of a domain can be very useful in understanding the potential of a specific technology particularly if one compares it to the improvement rate of competitive and complementary technological domains. This is because the improvement rates are reasonably consistent across time so a domain that is improving much more rapidly than a competitive domain will almost always eventually (even shortly) dominate the competitive markets.
By comparing the technological improvement rate of technological domains and sub-domains, we can better understand the
technology structure and see which technology develops faster and may dominate the market in the future.
Based upon prior discussion, relative rates of technical performance increase can have large implications for the future viability of component technologies in products and systems as well as the viability of industries and thus have great importance to forward-looking firms. Acquisition strategy, product component technology choice and appropriate research goals could be informed by improved understanding of the probable improvement potential of relevant technologies.

**METHODS AND PROCEDURES**

*Procedures*

My research is to explore a new technological domain and to decompose several sub-domains. I worked on patents from the Biochemical Pharmacology technological area and got 3 different domains and 4 of their sub-domains.
In the first period of my internship, I spent some time reading papers about patent search method then I applied this method to a test case to better understand how it works.
After that I chose a specific domain to do patent analyze by reference of Kay's technology map4 (Figure 1). Kay has defined 466 technology categories in his paper. The technology map is based on similarities in citing-to-cited relationships between categories of the International Patent Classification (IPC) of European Patent Office (EPO) patents from 2000 to 2006. In figure 1, each node color represents a technological area; lines represent relationships between technology categories (the darker the line the shorter the technological distance between categories;) labels for technological areas are placed close to the categories with largest number of patent applications in each area.

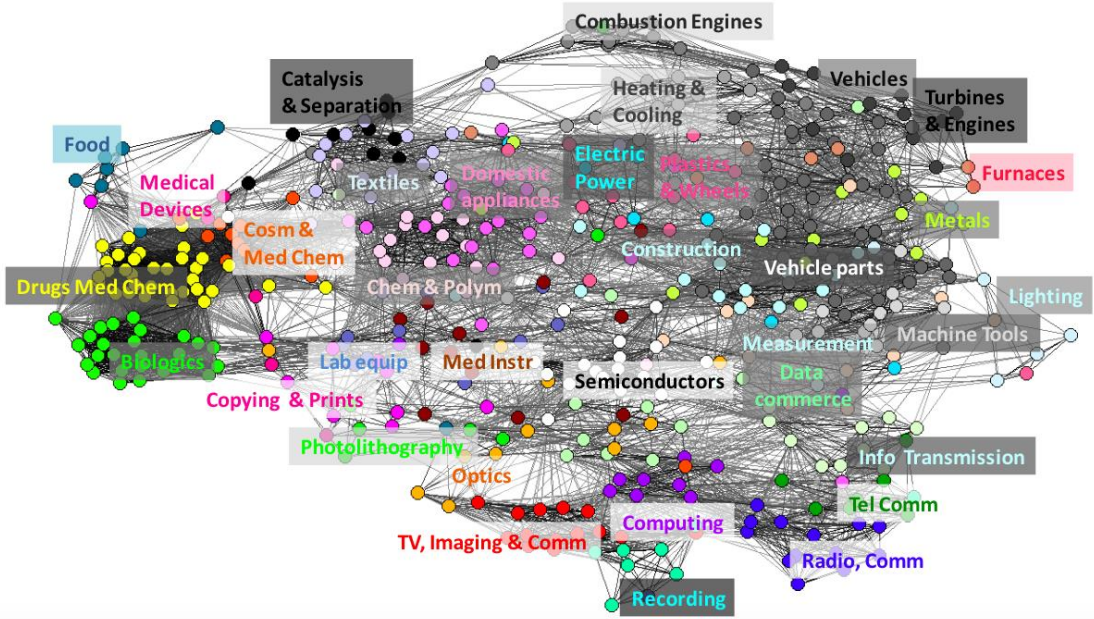

Figure 1. Full patent map of 466 technology categories and 35 technological areas

I start from a technological: Drugs, Medicines and Chemicals, because it seems they have very little overlaps with other technological domains. Also because the drug domain seems to develop very fast nowadays.

After choosing the technological area, I did some search to define the technological domains and their sub-domains and then used PatsNap to download all the patent data. Next I used RapidMiner to analyze those patent data and got the most relevant patent set. Then I applied the Classification Overlap Method (COM) to select the most relevant patent sets. Finally with the patent sets I can calculate the technology improvement rates of these technological domains. Data Source: Patents an PatsNap
In this research we use PatsNap to search and analyze patent information. Figure 213 shows the information for patent US7704946: Reversibles Inhibition of Pyramidal Gap Junction Activity, which we got from PatsNap. We can find all the standard information for a patent, such as: Title, Abstract, Publication Date, International Classification and US Classification. Those information are helpful in analyzing technological change.

Figure 2. Patent information

Patents consist abundant technology information thus make it possible to measure and compare technological improvements.

*Patent Classification Systems*

Our research are based mainly on US patents, donc we used two kinds of patent classifications to do our research, one is the international patent classification, the other is the United States patent classification.
The International Patent Classification (IPC) 11 is a hierarchical patent classification system used in over 100 countries to classify the content of patents in a uniform manner. Each classification term consists of a symbol such as A01B 1/00 (which represents "hand tools"). The first letter is the "section symbol" consisting of a letter from A ("Human Necessities") to H ("Electricity"). This is followed by a two digit number to give a "class symbol" (A01 represents "Agriculture; forestry; animal husbandry; trapping; fishing"). The final letter makes up the "subclass" (A01B represents "Soil working in agriculture or forestry; parts, details, or accessories of agricultural machines or implements, in general"). The subclass is then followed by a one-to-three-digit "group" number, an oblique stroke and a number of at least two digits representing a "main group" or "subgroup". A patent examiner assigns a classification to the patent application or other document at the most detailed level which is applicable to its contents.

• A: Human Necessities
• B: Performing Operations, Transporting
• C: Chemistry, Metallurgy
• D: Textiles, Paper
• E: Fixed Constructions
• F: Mechanical Engineering, Lighting, Heating, Weapons
• G: Physics
• H: Electricity

The United States Patent Classification12 is an official patent classification system used and maintained by the United States Patent and Trademark Office (USPTO). There are over 400 classes in the U.S. Patent Classification System, each having a title descriptive of its subject matter and each being identified by a class number. Each class is subdivided into a number of subclasses. Each subclass bears a descriptive title and is identified by a subclass number. The subclass number may be an integral number or may contain a decimal portion and/or alpha characters. A complete identification of a subclass requires both the class and subclass number and any alpha or decimal designations; e.g., 417/161.1A identifies Class 417, Subclass 161.1A.

*Previous Patent Search Method*

The most basic ways of searching for patents are the keyword search and the classification search. The keyword search uses search terms and Boolean operators (AND, OR, NOT, NEAR) to construct queries to find the most relevant patents. The classification method requires that the patents already be classified (such as in the US or International Patent classification systems), and that the patents in question can be pinpointed to just one or more patent classes.
The key words search use the occurrence of words which are sometimes very inaccurate especially when it comes to word sequences. As for the classification method, each patent may have several classifications and they may not have the same pertinence. So there is a need to use other patent search method to define the technology domain and find patents sets that can more accurately represent the technological domains.

*Classification Overlap method (COM)*

Beyond the two most basic methods for retrieving sets from the patent database, there have been an increasing number of approaches involving complex information retrieval techniques and methods. The method developed in Benson and Magee(2013) involves searching for keywords that are selected as potentially important in the technological domain of interest and analyzing the patents in each of the sets retrieved with the keyword analysis by quantitative metrics assessing the patent classes of the sets. The patents that are in both the most likely UPC patent class as well a the most likely IPC patent class are then taken as patents in the domain.

It is a relatively simple, objective and repeatable method for selecting sets of patents that are representative of a specific technological domain. The methodology consists of using search terms to locate the most representative international and
US patent classes and determines the overlap of those classes to arrive at the final set of patents. Comparison against traditional keyword searches and individual patent class searches shows that the classification overlap method can find a set of patents with more relevance and completeness and no more effort than the other two methods. Figure 3 represents the precess of this Classification Overlap Method.

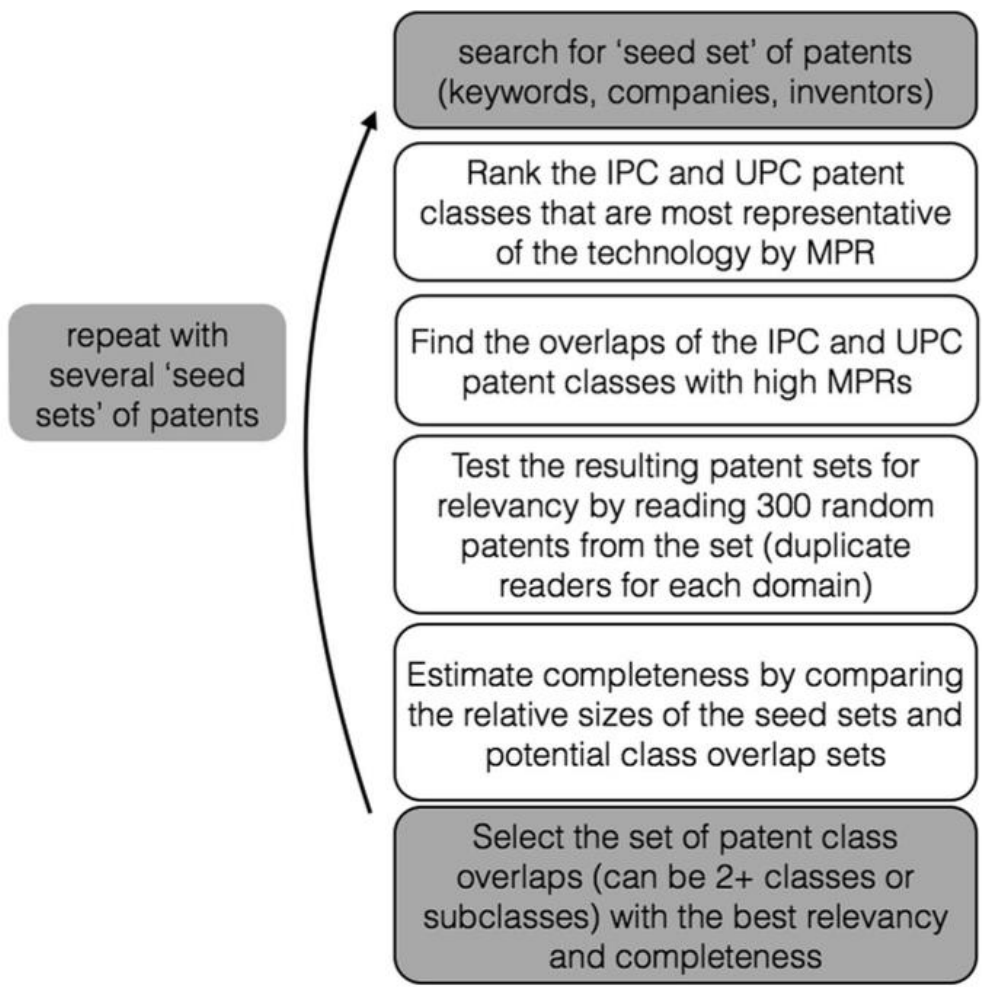

Figure 3. Process flow of COM

**Pre-search US issued patent titles, abstracts and claims for the search term**

The first step of the COM is to pre-search using a set of keywords to begin the process of finding the most representative patent classes. The input to the COM is simply a set of search terms (ex: Semiconductor) that can be entered into a text box. The pre-search was completed by searching for the word query in the title, abstract and claim of United States Issued Patents. Thus, the pre-search identifies a set of patents with specific query in the title or abstract or claim. The pre-search was done using the patent search tool PatsNap, which searched all US patents that were used as our database for further analysis.

**Rank the IPC and UPC patent classes that are most representative of the technology**

The next step in the COM is to use the set of patents resulting from the pre-search to determine the US patent classes (UPC) and international patent classes (IPC) that are most representative of the specific technology. The representativeness ranking for the patent classes is accomplished by using the mean-precision-recall (MPR) value. This value was inspired by the F1 score that is common in information retrieval, but uses the arithmetic mean (instead of the geometric mean) of the precision and recall of a returned data set. The recall for each of the listed patent classes is calculated by dividing the number of patents in pre-search results that are within the patent class by the number of patents in the pre-search patent set.

$$Recall = \frac{\#Patents\,in\,the\,Presearch\,within\,the\,PatentClass}{\#Retrived\,Patents\,in\,the\,Presearch}$$

Given the total size of the patent class, we determine the fraction of the patents in each patent class present in the pre-search, which is called the patent class precision. This normalizes the weight of very large and very small patent classes that may be over or under represented in the pre-search due to their different sizes. Calculate the precision of each patent class within the pre-search by dividing the number of patents in both the search and the the patent class by the total number of patents in the patent class.

$$Precision = \frac{\#Patents\,in\,the\,Presearch\,within\,the\,PatentClass}{\#Patents\,patentclass}$$

Finally we find the mean of precision and recall values, which gives us an estimate of how well each patent class represents the pre-search set. The MPR of each patent class is calculated by taking the mean of the patent class precision and patent class recall.

$$MPR = (Precision + Recall)/2$$

**Select the overlap of the most representative IPC class and UPC class**
To find the final set, the patents that are contained within both the IPC and UPC classes with the highest MPRs within the set of US issued patents are retrieved. The intuition for this step is founded upon the extensive examiner experience and knowledge embedded in these two classification systems. If a patent is listed in the most representative patent class in both systems (particularly since the two systems are somewhat differently structured), a reasonable hypothesis is that such dual membership results in obtaining patents of higher relevance.

Different technological domains have different overlap and may not have just one UPC and one IPC. So under certain circumstances we need to do multiple overlaps.
Figure 4. Different types of overlap types between multiple IPCs and UPCs using COM with specific sectors labeled. 5
As showed in figure 4, one technological domains may be selected to the overlaps of several UPCs and several IPCs.

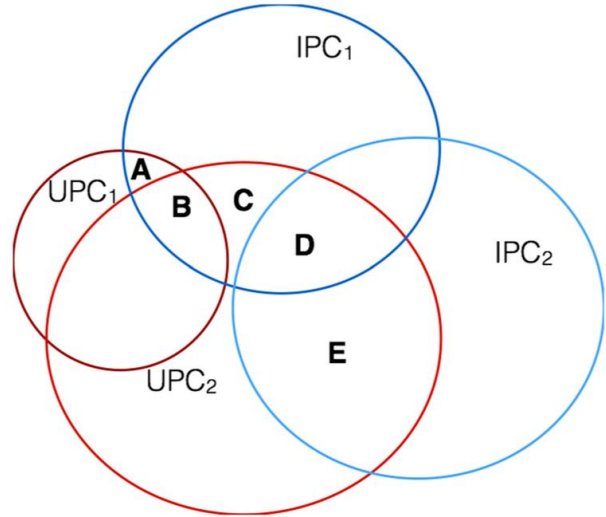

Figure 4. Different types of overlap types between multiple IPCs and UPCs using COM with specific sectors labeled[5].

*K value calculatio*n

After getting the most relevant patent sets, we can calculate the K value. First we need to calculate the patent indicators using the meta-data included in patents. The Regression model uses two indicators for calculating the estimated technological improvement rate: average publication year and average number of forward citations within 3 years of publication as described in Benson and Magee (2014c).

**Average Publication Year**

This is the average year of publication for patents within a technological domain. In this research, this includes all the patents that were listed in the database of PatsNap. This measure is calculated using equation:

$$\sum \frac{(t_i)}{SPC}$$

where SPC is the simple patent count: SPC = count(Pi). And ti is the publication year of patent *i*.

**Average number of Forward Citations within 3 years of Publication per patent**
This is the average number of forward citations that each patent received within 3 years of publication for patents in a technological domain. This measure is calculated using equation:

$$\sum \sum IF(t_{ij} - t_i) \leq 3$$

where $t_i$ is the publication year of patent $i$, $t_{ij}$ is the publication date of forward citation $j$ of patent $i$, and the function $IF(arg)$ only counts the values if the argument is satisfied.

The predictive model takes into account the average number of forward citations within 3 years and adds the average publication year of patents and is shown in equation:

$$K = -31.1285 + 0.0155 * AvePubYear + 0.1406 * Cite3$$

*Test study*

Table 1 shows an example of MPR calculation for the UPC and IPC in the pre-search for "Bipolar transistor".

Table 1. Example calculation of MPR for the search term Bipolar transistor

| Patent Class | Number of Patents in Pre-search and Patent Class | Patent Class Recall | Total number of Patents in Patent Class | Patent Class Precision | MPR |
|---|---|---|---|---|---|
| UPC: 257/370 | 495 | 0.027 | 689 | 0.718 | 0.372 |
| IPC: H01L29/73 | 3046 | 0.163 | 6818 | 0.447 | 0.305 |

The number of patents identified in the pre-search that are present in each class is shown in column two (this can be called the overlap of the pre-search and the patent class), it is found using the following search:

*TTL*: (Bipolar transistor) OR ABST: (Bipolar transistor) OR CLMS: (Bipolar transistor) AND DOCUMENT_TYPE: United States Issued Patent

The patent class Recall is shown in column 3. Column number 4 shows the total number of patents in each patent class, which is found by the following search:

*UPC: (257/370) AND DOCUMENT_TYPE: United States Issued Patent*
*IPC: (H01L29/73) AND DOCUMENT_TYPE: United States Issued Patent*

The patent precision is shown in column 5. The MPR of each patent class (column 6) is calculated by taking the mean of the patent class precision(column 5) and patent class recall(column 3). The most relevant patent set for Bipolar transistor is UPC – 257/370 & IPC - H01L29/73[6].

*Patent Search of 7 domains*

According to this classification system, I selected three technological domains: Drugs for Nervous system diseases treatment; Drugs for Cardiovascular diseases treatment and Drugs for Respiratory system diseases treatment.

The nervous system[8] is the part of an animal's body that coordinates its voluntary and involuntary actions and transmits signals to and from different parts of its body. Brain and nervous system problems are common. These neurological disorders include multiple sclerosis, Alzheimer's disease, Parkinson's disease, epilepsy, and stroke, and can affect memory and ability to perform daily activities. Then I chose two sub-domains for this technological domain: Drugs for Parkinson's and Drugs for Alzheimer's.

The essential components of the human cardiovascular system are the heart, blood and blood vessels. Cardiovascular disease (CVD) [9] is a class of diseases that involve the heart or blood vessels. Common CVDs include: ischemic heart disease (IHD), stroke, hypertensive heart disease, rheumatic heart disease (RHD), aortic aneurysms, cardiomyopathy, atrial fibrillation, congenital heart disease, endocarditis, and peripheral artery disease (PAD), among others. For this domain I chose Drugs for Hypertensive as its sub-domain.

The human respiratory system[10] is a series of organs responsible for taking in oxygen and expelling carbon dioxide. The primary organs of the respiratory system are lungs, which carry out this exchange of gases as we breathe. Diseases and onditions of the respiratory system fall into two categories: Viruses such as influenza, bacterial pneumonia and the new enterovirus respiratory virus that has been diagnosed in children; and chronic diseases, such as asthma and chronic obstructive pulmonary disease (COPD). For this domain I chose Drugs for Asthma as its sub-domain.

First I used the key word search for the seven domains I have chosen. The keyword search uses search terms and Boolean operators (AND, OR, NOT, NEAR) to construct queries to find the most relevant patents. I tried different combinations and found that just one or two keywords cannot get many patents. As there are a lot of patents in one domain it is more accurate to analyze more patents. So I used the most common disease names as the search term and the keyword drugs as we are interested in this specific domain.

**RESULTS**
*Final Patent Sets*

According to the rankings of IPCs and UPCs and by comparing the MPR value, table 2 shows the most relevant patent sets for 7 domains.

Table 2. Final Patent Sets

| Technology Domains | Pre-Search Patent Count | Patent Classes | Number of Patents in Pre-Search and Patent Classes | Total Patent Class Number | MPR |
|---|---|---|---|---|---|
| Drugs for Nervous System Diseases | 9902 | UPC: 424 OR 514 IPC: A61P25 | 307 | 16232 | 0.244 |
| Drugs for Alzheimer's | 3414 | UPC: 514 IPC: A61P25/28 | 984 | 5789 | 0.229 |
| Drugs for Parkinson's | 1065 | UPC: 424 OR 514 IPC: A61P25/16 | 452 | 2103 | 0.32 |
| Drugs for Cardiovascular System Diseases | 73524 | UPC: 424 OR 514 IPC: A61P9 | 5403 | 14358 | 0.225 |
| Drugs for Hypertensive | 3395 | UPC: 424 OR 514 IPC: A61P9/12 | 717 | 3937 | 0.2 |
| Drugs for Respiratory System Diseases | 9783 | UPC: 514 IPC: A61P11 | 2088 | 6071 | 0.214 |
| Drugs for Asthma | 3637 | UPC: 424 OR 514 IPC: A61P11/06 | 822 | 3137 | 0.244 |

In table 2, column 1 is the name of technology domains and sub-domains, column 2 is the number of patents we got by key word search, column 3 is the most relevant patent class we found by using the com, column 4 is the number of patents using both key word search and classification code search, column 5 is the patent number of technology domain. The last column is the MPR value calculated by equation:

(column4/column2 + column4/column5)/2

*Calculating the Patent Indicators and Using the Regression Model to Estimate Technological Improvement Rates*

After getting the most relevant patent sets, we can calculate the K value. First we need to calculate the patent indicators using the meta-data included in patents. The Regression model uses two indicators for calculating the estimated technological improvement rate: average publication year and average number of forward citations within 2 years of publication as described in Benson and Magee (2014c).

Table 3. Estimated K value

| Technology domains | AvePubYear | Cite3 | Predicted k |
|---|---|---|---|
| Drugs for Nervous System Diseases | 2000.28 | 2.36617 | 0.208583 |
| Drugs for Alzheimer's | 2002.82 | 2.35049 | 0.245692 |
| Drugs for Parkinson's | 2004.49 | 3.07129 | 0.372949 |
| Drugs for Cardiovascular System Diseases | 1999.91 | 2.51658 | 0.223867 |
| Drugs for Hypertensive | 1996.62 | 2.41072 | 0.158069 |
| Drugs for Respiratory System Diseases | 2001.53 | 2.79493 | 0.288165 |
| Drugs for Asthma | 2005.14 | 2.99171 | 0.371769 |

After getting the k values for the 7 domains and sub-domains, I compared them to the k value of some well-known domains that Prof. Chris and Benson has calculated before. The result is shown in Figure 5.

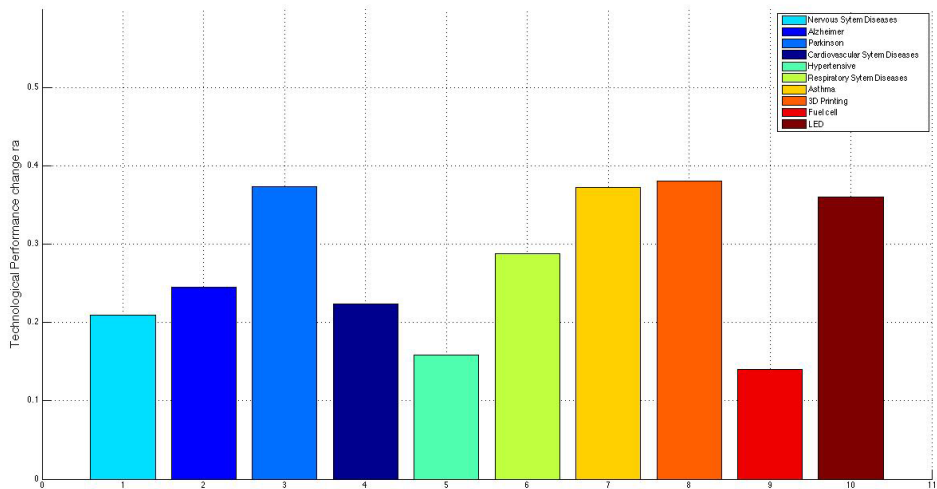

Figure 5. Technological performance improvement rate

## DISCUSSION

*Implications for technology strategy for firms*

The technological improvement rate of a domain can be very useful in understanding the potential of a specific technology particularly if one compares it to the improvement rate of competitive and complementary technological domains. This is because the improvement rates are reasonably consistent across time so a domain that is improving much rapidly than a competitive domain will almost always eventually (even shortly) dominate the competitive markets (except for a few resistantniches). Thus, quantitative technology improvement rates are helpful in understanding the future of technology from the component level to entire industries.

As technological domain is complicated, we tried in this research to decompose technological domains to sub-domains. We found out that the performance improvement rates in the Biochemical Pharmacology technological area are different. Compared to several technologies like LED, Batteries and 3D-Printing, the pharmaceutical area is developing very fast. Some diseases like Parkinson's and Asthma develop faster than other sub-domains.

Relative rates of technological performance increase can have large implications for the future viability of component technologies in products and systems as well as viability of industries and thus have great importance for forward-looking firms. Acquisition strategy, product component technology choice and appropriate research goals could be informed by improved understanding of the probable improvement potential of relevant technologies. Moreover, the results of performance improvement monitoring have implications for choosing technologies that should receive research funding from firms and governments and for choosing ventures in which to invest risk capital.

*Limitations and Future Research*

Limitations of the current study and further useful work includes continued improvement of the COM and continued use of the method to further explore overall technological structure.

While the method for assessing relevancy (dual readers of all patents with resolution by three participants when rare divergences appear) is effective, it is time consuming and the most non automated and potentially subjective part of the COM.

Future work also includes research in non overlap patents between technological sub-domains and see how they influence the performance change rate of technologies.

**CONCLUSIONS**

This report represents the patent search of 3 technological domains and 4 sub-domains in the Biochemical Pharmacology technological area using the Classification Overlap Method (COM) and the calculating of their technological improvement rates. The method used for executing the patent search is an extended version of a method previously described (Benson and Magee 2013). The extension involves more emphasis upon multiple IPC and UPC class listings to be utilized in the gathering of the final patent set and a deeper level of technological domain. As in the earlier work in our group, the effectiveness of this method indicates that the US Patent examiners are using the two classification systems differently enough to make joint groups of patents more aligned with (relevant to) the technological domains than patent sets using a singular classification system. Over several technological domains within the Biochemical Pharmacology technological area, the

COM is shown to have highly relevant sets of patents where relevance is empirically assessed by reading of patents. The COM is also shown here to give a fairly complete set of patents as assessed by use of multiple seed patent sets and analysis of all of the resulting possible overlaps.

This research shows a relatively high technological improvement rate in the Biochemical Pharmacology technological area and gives us a better understanding of technological structure within this area. Different technological domains have a high patent overlap as they share common classification code.

Technological sub-domains are included in their technological domains but their technological improvement rates are different. Some domains like Drugs for Parkinson's or Drugs for Asthma have higher performance improvement rates.

**ACKNOWLEDGEMENTS**
I would like to express my deepest appreciation to professor Christopher L. Magee who provided me the possibility to work in the International Design Center where I have learned a lot through literature meetings and articles.

I would also like to acknowledge with much appreciation the crucial role of Dr. Hyunseok Park, who helped me a lot by sharing those creative thoughts with me.

A special thanks to Arlyn Hertz and Mayoka Takemori who helped me so much with my Visa and my status at MIT.

ABOUT THE STUDENT AUTHOR
Xin Guo is actually a graduate student in ENSTA ParisTech, France and she's also pursuing her master's degree in the university of Paris-Saclay. She will graduate as an engineer in the year of 2016 after finishing her end-of-year internship. This paper is about her work as an intern in the International Design Center in MIT during May and August 2015.